\documentclass[aps,prx,twocolumn,showpacs,superscriptaddress,longbibliography]{revtex4-2}

\usepackage{amsmath,amssymb,graphicx}
\usepackage{graphicx,epstopdf}
\usepackage{gensymb}
\usepackage{upgreek}
\usepackage{float}
\usepackage[dvipsnames]{xcolor}
\usepackage[utf8]{inputenc}
\begin{document}
\title{Anisotropic collapse of electronic correlations in the ferromagnet UGe$_2$\\ under high magnetic field}

\author{K. Somesh}
\affiliation{Laboratoire National des Champs Magnétiques Intenses- EMFL, CNRS, Univ. Grenoble Alpes, INSA-T, Univ. Toulouse 3, 31400 Toulouse, France}
\author{T. Thebault}
\affiliation{Laboratoire National des Champs Magnétiques Intenses- EMFL, CNRS, Univ. Grenoble Alpes, INSA-T, Univ. Toulouse 3, 31400 Toulouse, France}
\affiliation{Institute for Quantum Materials and Technologies, Karlsruhe Institute of Technology, D-76021 Karlsruhe, Germany}
\author{V. Taufour}
\affiliation{Univ. Grenoble Alpes, CEA, Grenoble INP, IRIG, PHELIQS, 38000, Grenoble, France}
\affiliation{Department of Physics and Astronomy, University of California, Davis, California 95616, USA}
\author{D. Aoki}
\affiliation{Institute for Materials Research, Tohoku University, Ikaraki 311-1313, Japan}
\author{F. Duc}
\affiliation{Laboratoire National des Champs Magnétiques Intenses- EMFL, CNRS, Univ. Grenoble Alpes, INSA-T, Univ. Toulouse 3, 31400 Toulouse, France}
\author{G. Knebel}
\affiliation{Univ. Grenoble Alpes, CEA, Grenoble INP, IRIG, PHELIQS, 38000, Grenoble, France}
\author{D. Braithwaite}
\affiliation{Univ. Grenoble Alpes, CEA, Grenoble INP, IRIG, PHELIQS, 38000, Grenoble, France}
\author{W. Knafo}
\affiliation{Laboratoire National des Champs Magnétiques Intenses- EMFL, CNRS, Univ. Grenoble Alpes, INSA-T, Univ. Toulouse 3, 31400 Toulouse, France}

\date{\today}

\begin{abstract}
We present electrical-resistivity measurements on the prototypical heavy-fermion ferromagnet UGe$_2$ under pulsed magnetic field up to 60~T. An anisotropic field-induced suppression of the electronic correlations is revealed. The electrical resistivity strongly decreases when a magnetic field $\mathbf{H}$ is applied along the easy magnetic axis $\mathbf{a}$, while it remains almost unchanged when $\mathbf{H}$ is applied along the hard magnetic axes $\mathbf{b}$ and $\mathbf{c}$. The field-induced destabilization of the ferromagnetic state is also anisotropic: the anomaly at the Curie temperature $T_C$ disappears in fields higher than $\gtrsim1$~T for $\mathbf{H}\parallel\mathbf{a}$ and in fields higher than $\gtrsim20$~T for $\mathbf{H}\parallel\mathbf{b},\mathbf{c}$. At temperatures below 2~K, we observe quantum oscillations in fields larger than 50~T applied along $\mathbf{b}$, which support the presence of a two-dimensional Fermi surface similar to that previously observed at low fields.
\end{abstract}
\maketitle

\section{Introduction}
\label{Intro}

The interest for U-based ferromagnets \cite{Aoki2019a} was recently renewed by the proposition that the superconductor UTe$_2$ may be a nearly ferromagnet, i.e., a paramagnet in which ferromagnetic (FM) fluctuations would be responsible for triplet superconductivity \cite{Ran2019a}. Magnetic-field-induced or reinforced superconductivity was first reported in the ferromagnets UGe$_2$ \cite{Saxena2000,Sheikin2001}, URhGe \cite{Aoki2001,Levy2005}, and UCoGe \cite{Huy2007,Aoki2009,Wu2017} before being discovered in UTe$_2$, which was suggested to be a paramagnetic end-compound of this family of U-based ferromagnetic superconductors \cite{Ran2019a,Ran2019b,Knebel2019,Knafo2021a,Aoki2021,Valiska2021,Wu2025,Lewin2024a,Frank2024}. However, neutron scattering experiments later proved that UTe$_2$ is a nearly antiferromagnet, with the observation of antiferromagnetic (AF) fluctuations at ambient pressure \cite{Duan2020,Knafo2021b,Butch2022} and long-range AF order under pressure \cite{Knafo2025}. This highlighted the difficulty in determining the FM or AF nature of the electronic correlations using non-$\mathbf{k}$-resolved experimental probes, as electrical-transport, thermodynamic, nuclear-magnetic-resonance (NMR) and muon-relaxation measurements (see for instance \cite{Sundar2019,Lin2020,Thomas2020,Aoki2020,Valiska2021,Willa2021,Tokunaga2022,Azari2023}). In addition, while electrical resistivity is widely used to investigate the magnetic and superconducting properties of metallic systems under intense pulsed magnetic fields \cite{Knafo2021c}, we noticed that a systematic study of a reference U-based ferromagnet under wide ranges of temperatures and intense magnetic fields of different directions has not been performed so far, limiting, at least partly, our ability to recognize the signatures of ferromagnetic correlations using this technique. This motivated us to perform the electrical-resistivity study of a ferromagnet under intense magnetic fields presented here.

The above-mentioned compound UGe$_2$ can be considered as a textbook U-based ferromagnet \cite{Aoki2019a}. It orders ferromagnetically at temperatures below the Curie temperature $T_C\simeq 52$~K and its ferromagnetically-ordered moment reaches $\mu_m \simeq 1.4~\mu_{B}/\mathrm{U}$ at low temperature \cite{Boulet1997,Kernavanois2001}. Ferromagnetic ordering at $T_C$ leads to step-like anomalies in the heat capacity $C_p$, thermal expansion $\alpha$, and in the electrical resistivity $\rho$ \cite{Huxley2001,Hardy2009,Troc2012}. Within the ferromagnetic phase, electronic correlations also drive broad extrema at the temperature $T^{*}\simeq 30$~K in the electronic heat capacity divided by temperature $C^e_p/T$ (estimated after phonon subtraction) \cite{Raymond2006,Hardy2009}, thermal expansion $\alpha$ \cite{Hardy2009} and in the temperature derivative of the electrical resistivity $\partial\rho/\partial T$ \cite{Bauer2001,Troc2012}. While the magnetization $M$ shows a smooth variation through the crossover at $T^*$, its temperature-derivative $\partial M/\partial T$ presents a broad extremum at $T^*$ \cite{Hardy2009,Troc2012}. Magnetic-susceptibility measurements showed that UGe$_2$ has an Ising magnetic anisotropy, $\mathbf{a}$ being the easy magnetic axis while $\mathbf{b}$ and $\mathbf{c}$ are hard magnetic axes \cite{Galatanu2005}. A magnetization study performed at the temperature $T \simeq 4.2$~K, i.e., well below $T_C$, confirmed that the ferromagnetic moments are aligned along $\mathbf{a}$ and showed no trace of metamagnetism beyond the low-field alignment of the ferromagnetic moments, under magnetic fields $\mu_{0}\mathbf{H} \parallel\mathbf{a},\mathbf{b}$ up to 27~T and $\mu_{0}\mathbf{H}\parallel\mathbf{c}$ up to 14~T \cite{Sakon2007}. Resistivity studies under magnetic fields up to $\mu_{0}H = 15$~T for $\mathbf{H}\parallel\mathbf{a}$, $\mathbf{b}$, and $\mathbf{c}$ and along intermediate directions of field, were interpreted as the signature of open orbits along $\mathbf{b}$ and $\mathbf{c}$ \cite{Onuki1991,Satoh1992_1}. De-Haas-van-Alphen and Shubnikov-de-Haas quantum-oscillations experiments under magnetic fields up to 15~T showed a Fermi surface made of several bands, the main ones being nearly cylindrical along $\mathbf{b}$ \cite{Onuki1991,Satoh1992_2,Settai2002,Terashima2002,Palacio2016}.

\begin{figure*}[ht]
	\includegraphics[width=\textwidth]{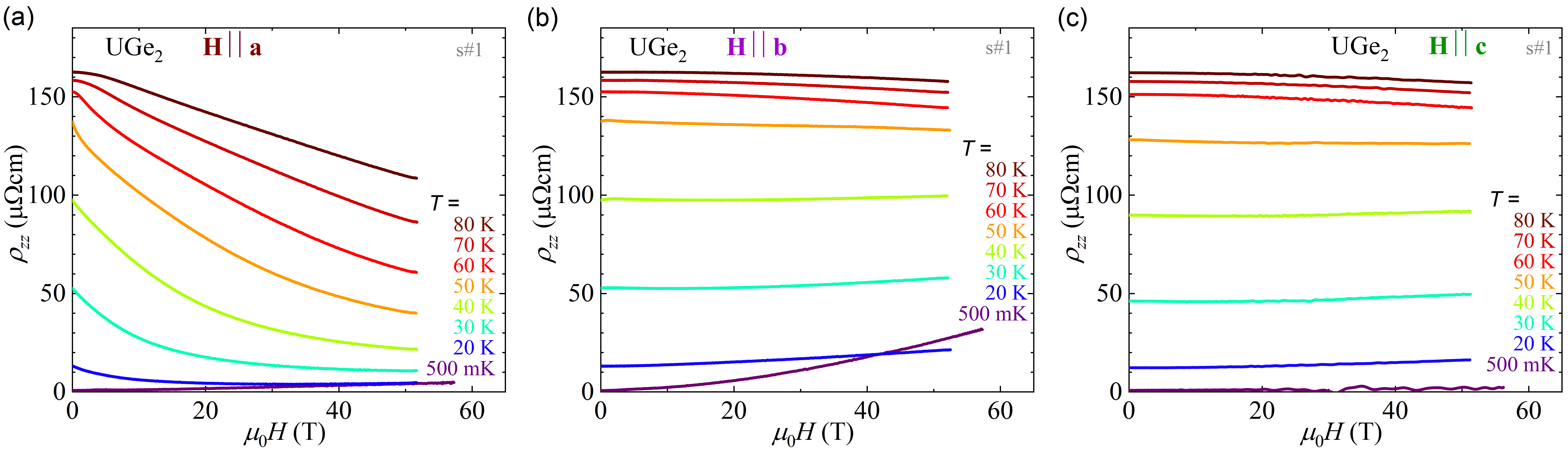}
	\caption{\label{Fig1} Magnetic-field dependence of the electrical resistivity $\rho_{zz}$ of UGe$_2$ at temperatures from 500~mK to 80~K under magnetic fields (a) $\mathbf{H}\parallel\mathbf{a}$, (b) $\mathbf{H}\parallel\mathbf{b}$, and (c) $\mathbf{H}\parallel\mathbf{c}$.}
\end{figure*}

Under pressure $p$, the Curie temperature $T_C$ decreases continuously before collapsing at the critical pressure $p_c=1.6$~GPa, above which a paramagnetic ground state is reached \cite{Takahashi1993,Saxena2000}. For $p\gtrsim0.8$~GPa, the crossover at the temperature $T^*$ is replaced by a first-order phase transition at the temperature $T_x\lesssim20$~K associated with step-like variations in the magnetization $M$ \cite{Tateiwa2001a,Pfleiderer2002} and in the electrical-resistivity temperature derivative $\partial\rho/\partial T$ \cite{Bauer2001}. A transition from a high-temperature ferromagnetic phase FM1 with a small ordered moment to a low-temperature phase FM2 with a large ordered moment occurs at $T_x$ \cite{Pfleiderer2002}. $T_x$ decreases under pressure and collapses at the pressure $p_x=1.2$~GPa \cite{Oomi1995,Bauer2001,Pfleiderer2002}. A superconducting phase was reported within the ferromagnetic phase, with a maximum of the superconducting temperature $T_{sc}$ in the vicinity of $p_x$, and was suspected to be mediated by a triplet mechanism \cite{Saxena2000,Tateiwa2001b,Taufour2010}. The enhancement of the electrical-resistivity Fermi-liquid coefficient $A$ close to $p_x$ \cite{Tateiwa2001b} also indicates the presence of critical magnetic fluctuations, which may play a role for superconductivity. A Fermi-surface reconstruction was reported at $p_x$ \cite{Settai2002,Terashima2002}, indicating an interplay between the Fermi surface, the magnetic fluctuations and superconductivity. The combination of pressure and magnetic fields applied along $\mathbf{a}$ also showed that beyond $p_c$ a metamagnetic transition is induced, driving a wing-like three-dimensional phase diagram associated with a quantum critical end-point \cite{Taufour2010}. Superconductivity was further found to be reinforced in the vicinity of the pressure-induced metamagnetic field, supporting that the high-field magnetic and superconducting properties are intimately related \cite{Sheikin2001}.

Although UGe$_2$ has been intensively studied since three decades \cite{Aoki2019a}, investigations of this compound under magnetic fields beyond 20~T remain scarce \cite{Sakon2007}. In this work, we revisit the high-magnetic field electrical resistivity of UGe$_2$ with the objective to characterize the signatures of the electronic correlations in this prototypical U-based ferromagnet. We present a study under a large range of temperatures from 500~mK to 80~K and magnetic fields up to 60~T applied along the easy magnetic axis $\mathbf{a}$ and the hard magnetic axes $\mathbf{b}$ and $\mathbf{c}$ (complementary data under magnetic fields along directions tilted from $\mathbf{b}$ toward $\mathbf{a}$ or $\mathbf{c}$ are shown in the Supplemental Material \cite{SM}). We observe an anisotropic resistivity collapse in field, which is related with the anisotropy of the magnetic properties. We also report quantum oscillations under magnetic fields larger than 50~T applied along $\mathbf{b}$, which confirm the presence of a cylindrical Fermi surface.

\section{Experimental details}
\label{Exp}

The UGe$_2$ single crystal investigated here was synthesized by the Czochralski pulling technique \cite{Huxley2001,Taufour2010}. The sample, of dimensions of $\simeq0.4$~mm along $\mathbf{a}$, $\simeq0.2$~mm along $\mathbf{b}$, and $\simeq1.5$~mm along $\mathbf{c}$, was cut by electrical spark. Its crystallographic orientation was confirmed at room temperature using a Laue diffractometer. Electrical contacts were made by spot welding of 15-$\mathrm{\upmu m}$ gold wires. The electrical resistivity $\rho_{zz}$ was measured by the four-contact method, with an electrical current $\mathbf{I}\parallel\mathbf{c}$ of 10~mA and a frequency $f\approx60$~kHz, in a 60-T pulsed magnet at the Laboratoire National des Champs Magnétiques Intenses (LNCMI) in Toulouse. The sample geometrical factor could not been determined directly from its shape and our resistivity data have been normalized so that $\rho_{zz}(T_C)=145$~$\mu\Omega$cm, in agreement with previously published data \cite{Troc2012}. Magnetic-field pulses of 50-ms-rise (up sweep) and 280-ms-fall (down sweep) have been produced using 6-MJ and 14-MJ generators. The data presented here were collected during the down sweeps. Measurements have been made under magnetic fields $\mathbf{H}\parallel\mathbf{a}$, $\mathbf{b}$ and $\mathbf{c}$, at constant temperatures $T$ from 500~mK to 80~K using an $^3$He insert combined with a standard $^4$He cryostat. A single-axis rotator was used, allowing complementary measurements at different angles $\theta = \rm\left(\mathbf{H},\mathbf{b}\right)$ (with $\mathbf{H}\perp\mathbf{c}$) and $\varphi = \rm\left(\mathbf{H},\mathbf{b}\right)$ (with $\mathbf{H}\perp\mathbf{a}$), which are presented in the Supplemental Material \cite{SM}.

\begin{figure*}[t]
	\includegraphics[width=\textwidth]{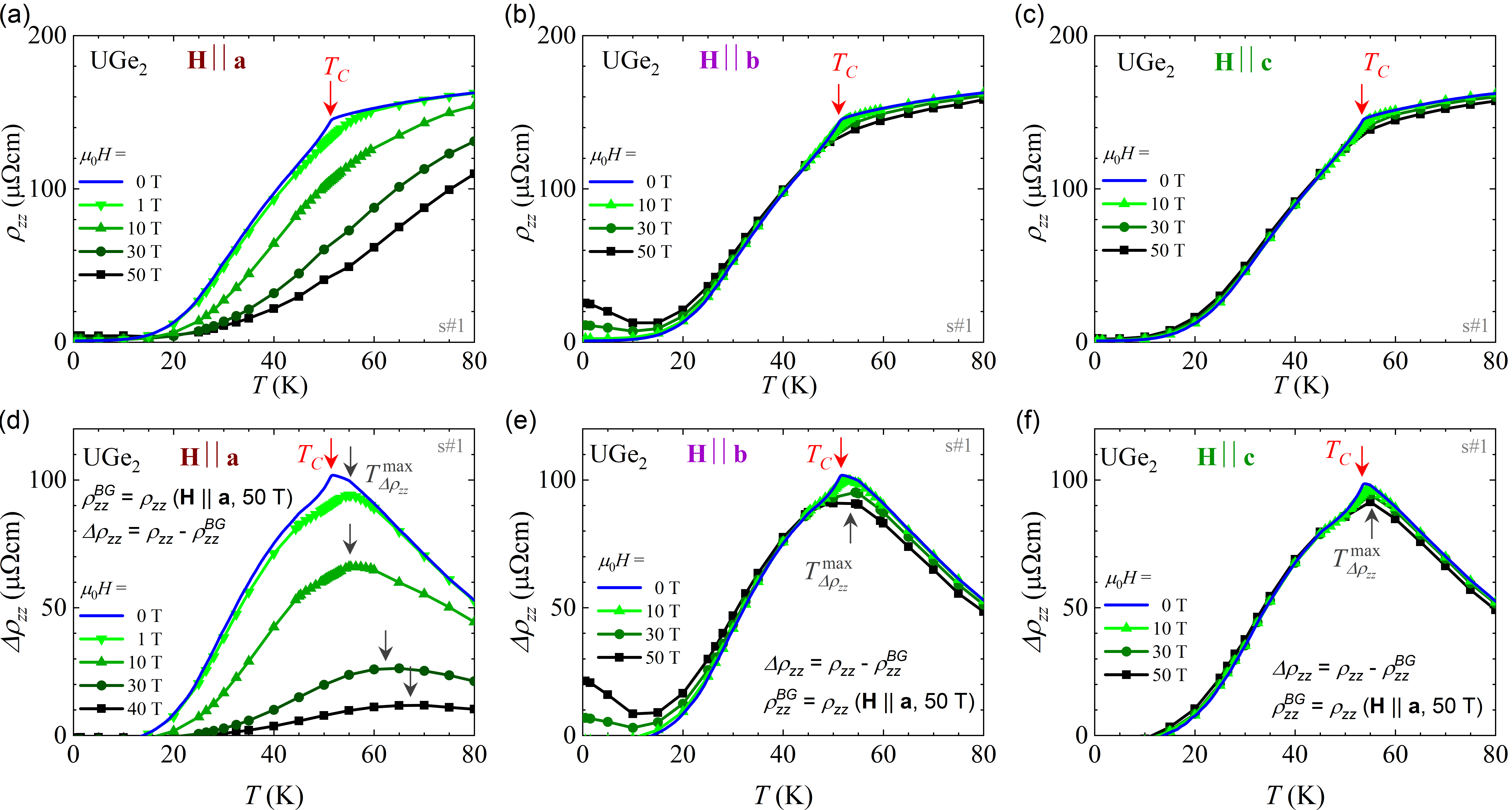}
	\caption{\label{Fig2} Temperature dependence of $\rho_{zz}$ of UGe$_2$ at constant magnetic fields (a) $\mu_0\mathbf{H}\parallel\mathbf{a}$, (b) $\mu_0\mathbf{H}\parallel\mathbf{b}$, and (c) $\mu_0\mathbf{H}\parallel\mathbf{c}$ up to 50~T. Temperature dependence of $\Delta\rho_{zz}$, estimated after subtraction of a background resistivity $\rho_{zz}^{BG} = \rho_{zz}(\mathbf{H}\parallel\mathbf{a}, \rm{50~T})$, at constant magnetic fields (d) $\mu_0\mathbf{H}\parallel\mathbf{a}$, (e) $\mu_0\mathbf{H}\parallel\mathbf{b}$, and (f) $\mu_0\mathbf{H}\parallel\mathbf{c}$ up to 50~T.}
\end{figure*}

\section{Results}
\label{Results}

Figures \ref{Fig1}(a-c) present the electrical resistivity $\rho_{zz}$ of UGe$_2$ measured as a function of magnetic field up to 60~T, with $\mathbf{H}\parallel\mathbf{a}$, $\mathbf{b}$, and $\mathbf{c}$, respectively, and for different temperatures from 500~mK to 80~K.
Anisotropic behaviors are revealed, depending on the direction of the magnetic field applied along or perpendicular to the easy magnetic axis.
Figure \ref{Fig1}(a) shows that, for $\mathbf{H}\parallel\mathbf{a}$, which is the easy-magnetic axis, $\rho_{zz}$ decreases continuously with $H$ at all temperatures larger than 20~K. The field variation of $\rho_{zz}$ is the strongest at $T\simeq 50$~K, i.e., very near to $T_C$, where $\rho_{zz} (0~{\rm T}) - \rho_{zz} (50~{\rm T})\simeq 90~\mu\Omega\rm{cm}$ is $\simeq9$ times larger than at $T=20$~K (the corresponding strongest field-variation in normalized values $[\rho_{zz}-\rho_{zz}(H=0)]/\rho_{zz}(H=0)$ occurs at $\simeq30$~K, see Figure S1 in the Supplemental Material \cite{SM}). A change of curvature in the field variation of $\rho_{zz}$ is visible near to $T_C$ at small magnetic fields $\mu_{0}H \lesssim 2$~T (see Figure S3(a) in the Supplemental Material \cite{SM}). At the lowest temperatures, a slow increase of $\rho_{zz}$ with $H$ is visible and may be related with a cyclotron motion of carriers induced by the magnetic field. Figure \ref{Fig1}(b) shows that, for $\mathbf{H}\parallel\mathbf{b}$, which is a hard magnetic axis, $\rho_{zz}$ remains almost unchanged under magnetic field at all temperatures $T\geq15$~K. At low temperatures, $\rho_{zz}$ strongly increases with the magnetic field (see Figures S1 and S4 in the Supplemental Material \cite{SM}), which is attributed to a field-induced cyclotron motion of the carriers. Figure \ref{Fig1}(c) shows that, for $\mathbf{H}\parallel\mathbf{c}$, which is a hard magnetic axis too, $\rho_{zz}$ is almost constant over the measured field range at all temperatures.

 \begin{figure*}[t]
  	\includegraphics[width=\textwidth]{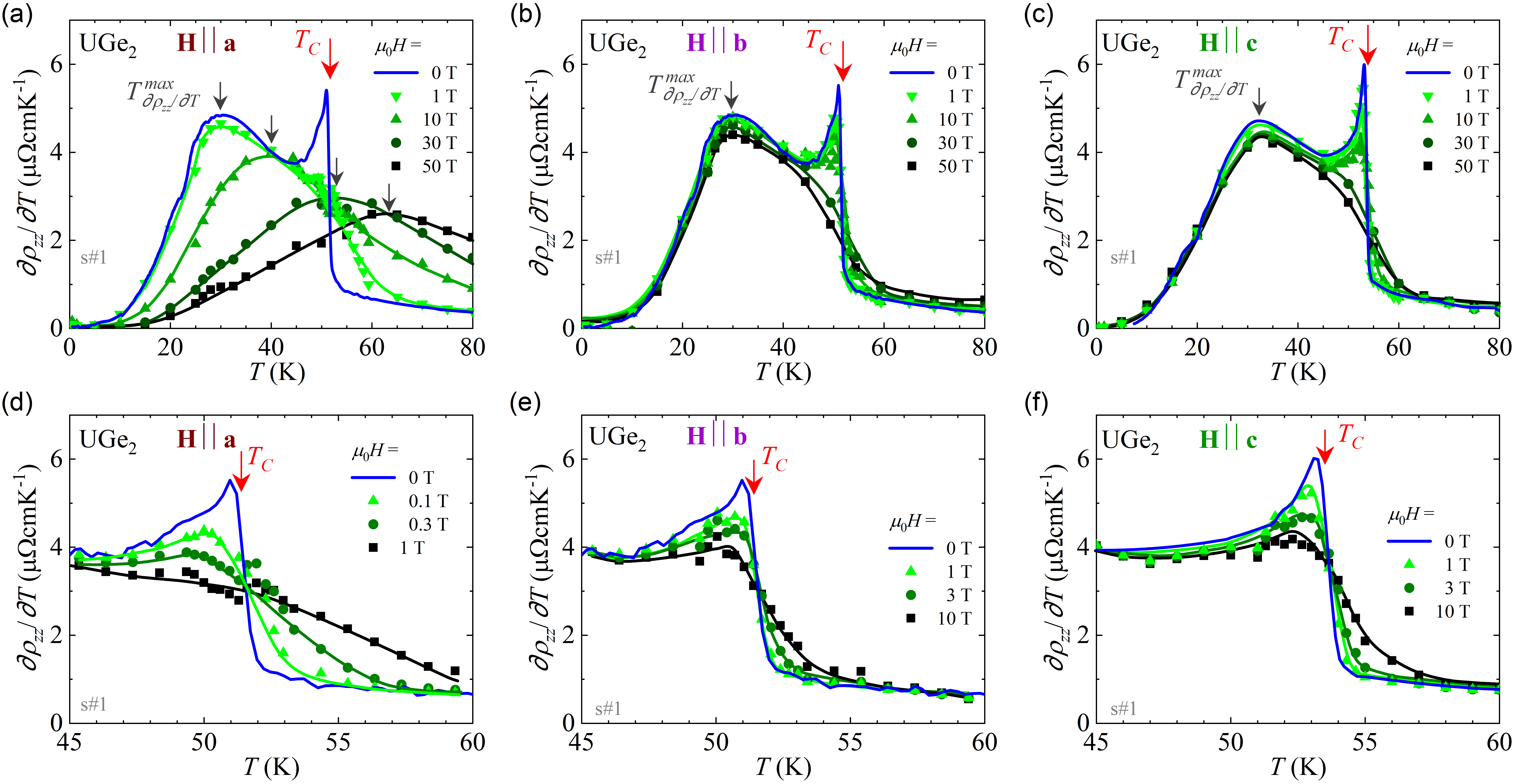}
  	\caption{\label{Fig3} Temperature dependance of $\partial\rho_{zz}/\partial T$ of UGe$_2$  at constant magnetic fields (a) $\mu_0\mathbf{H}\parallel\mathbf{a}$, (b) $\mu_0\mathbf{H}\parallel\mathbf{b}$, and (c) $\mu_0\mathbf{H}\parallel\mathbf{c}$ up to 50~T. $\partial\rho_{zz}/\partial T$ versus $T$ in reduced temperature and magnetic field windows, with $T$ from 45 to 60~K and (d) $\mu_0\mathbf{H}\parallel\mathbf{a}$ up to 1~T, (e) $\mu_0\mathbf{H}\parallel\mathbf{b}$ and (f) $\mu_0\mathbf{H}\parallel\mathbf{c}$ up to 10~T.}
  \end{figure*}

Figures \ref{Fig2}(a-c) show $\rho_{zz}$ versus $T$ data at constant magnetic fields $\mu_{0}H$ from 0 to 50~T applied along $\mathbf{a}$, $\mathbf{b}$, and $\mathbf{c}$, extracted from the $\rho_{zz}$ versus $H$ data presented in Figures \ref{Fig1}(a-c), respectively. The zero-field data are in agreement with previously-published data \cite{Onuki1992,Bauer2001}. $\rho_{zz}$ presents a broad shoulder at temperatures between 20 and 80~K, indicating the onset of electronic correlations. $\rho_{zz}$ also shows a kink at the Curie temperature $T_C\simeq52$~K, which marks the onset of the low-temperature ferromagnetic phase. Figure \ref{Fig2}(a) confirms that $\rho_{zz}$ strongly decreases with increasing magnetic fields for $\mathbf{H}\parallel\mathbf{a}$, whereas Figures \ref{Fig2}(b,c) show that $\rho_{zz}$ is almost field-invariant for $\mathbf{H}\parallel\mathbf{b}$ and  $\mathbf{H}\parallel\mathbf{c}$. Figure \ref{Fig2}(b) shows that, for $\mathbf{H}\parallel\mathbf{b}$, the large high-magnetic-field increase of $\rho_{zz}$ ascribed to a cyclotron motion of carriers is visible at temperatures $T\leq15$~K. For the three magnetic-field directions, the kink visible in $\rho_{zz}$ at $T_C\simeq 52$~K and zero field is rapidly suppressed in field and is replaced by a smooth variation of $\rho_{zz}$. A focus on the field-induced suppression of the anomaly at $T_C$ is made in the following.

Figures \ref{Fig2}(d-f) show the temperature dependence of $\Delta\rho_{zz}=\rho_{zz}-\rho_{zz}^{BG}$ determined for $\mathbf{H}\parallel\mathbf{a}$, $\mathbf{b}$, and $\mathbf{c}$, respectively. $\Delta\rho_{zz}$ is an estimation of the contribution to the electrical resistivity driven by the low-temperature electronic correlations. Here, assuming that most of the low-temperature electronic correlations have vanished in a magnetic field $\mu_0\mathbf{H}\parallel\mathbf{a}$ of 50~T, which corresponds to a regime deep inside the polarized paramagnetic regime (PPM) established beyond a magnetic field of a few hundreds of mT \cite{Lhotel2003}, we made the approximation that a 'background' resistivity can be estimated as $\rho_{zz}^{BG}=\rho_{zz}(\mathbf{H}\parallel\mathbf{a},50~\rm{T})$. Similar approximations were recently made for electrical-resistivity studies of UTe$_2$ \cite{Valiska2021,Thebault2022,Thebault2024}. The plot of $\Delta\rho_{zz}$ versus $T$ permits to emphasize the presence of several contributions. In zero magnetic field, a broad maximum of $\Delta\rho_{zz}$ centered around $\simeq50$~K, with a full width at half maximum $\Delta T\simeq50$~K, is the signature of electronic correlations in the system. On top of this broad signal, a sharp kink at $T_C\simeq52$~K indicates the onset of ferromagnetism and coincides with the maximal value of $\Delta\rho_{zz}$. Figure \ref{Fig2}(d) shows that, in a magnetic field applied along the easy magnetic axis $\mathbf{a}$, a fast decrease of $\Delta\rho_{zz}$ indicates a collapse of the electronic correlations. In fields $\mu_0H\geq1$~T, the kink at $T_C$ has disappeared and the temperature at the broad maximum of $\Delta\rho_{zz}$ increases with increasing field, up to $T_{\Delta\rho_{zz}}^{max}\simeq67$~K at $\mu_0H=40$~T. Figures \ref{Fig2}(e-f) show that, in magnetic fields applied along the hard magnetic axes $\mathbf{b}$ and $\mathbf{c}$, respectively, $\Delta\rho_{zz}$ has slower variations than for $\mathbf{H}\parallel\mathbf{a}$. The kink at $T_C$ disappears in fields $\gtrsim20$~T and the broad anomaly centered around $T_{\Delta\rho_{zz}}^{max}\simeq50$~K is almost unaffected under magnetic fields up to 50~T. In Figure \ref{Fig2}(e), the cyclotron-motion effect induced at low temperature is also visible under the largest fields applied along $\mathbf{b}$.

\begin{figure*}[t]
	\includegraphics[width=\textwidth]{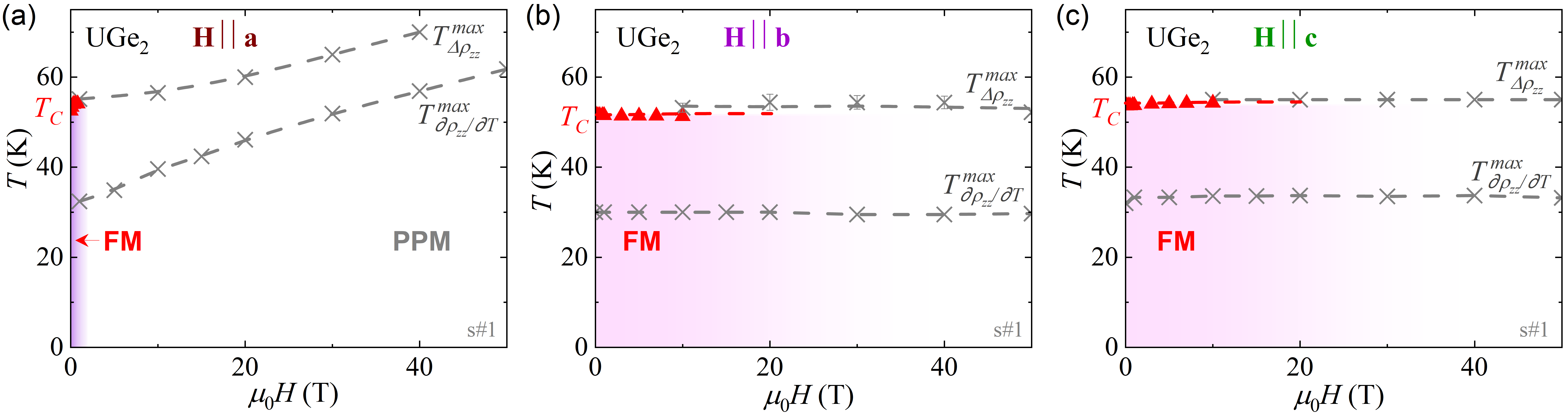}
	\caption{\label{Fig4} Magnetic-field-temperature phase diagram of UGe$_2$ under a magnetic field (a) $\mathbf{H}\parallel\mathbf{a}$, (b) $\mathbf{H}\parallel\mathbf{b}$, and (c) $\mathbf{H}\parallel\mathbf{c}$. FM and PPM label the ferromagnetic phase and the polarized paramagnetic regime, respectively.}
\end{figure*}

Figures \ref{Fig3}(a-c) present the temperature derivative of the electrical resistivity $\partial\rho_{zz}/\partial T$ versus $T$ extracted from our data for $\mathbf{H}\parallel\mathbf{a}$, $\mathbf{b}$, and $\mathbf{c}$, respectively. At zero magnetic field, $\partial\rho_{zz}/\partial T$ shows a broad maximum centered at the temperature $T_{\partial\rho_{zz}/\partial T}^{max} \simeq 30$~K. This anomaly and the broad maximum in $\Delta\rho_{zz}$ at the temperature $T_{\Delta\rho_{zz}}^{max}\simeq50$~K result from the same crossover driven by electronic correlations (see Discussion in Section \ref{Discussion}). In addition to the broad anomaly, a sharp step-like feature in $\partial\rho_{zz}/\partial T$ is visible at $T_C\simeq52$~K. For $\mathbf{H}\parallel\mathbf{a}$, Figure \ref{Fig3}(a) shows that the amplitude of the broad anomaly decreases with increasing field, from a maximal value of $\simeq4.5~\mu\Omega$ cm K$^{-1}$ at zero field to $\simeq2.5~\mu\Omega$ cm K$^{-1}$ at $\mu_0H=50$~T, while the associated temperature  $T_{\partial\rho_{zz}/\partial T}^{max}$ increases up to 65~K at $\mu_0H=50$~T. A focus presented in Figure \ref{Fig3}(d) shows that, for $\mathbf{H}\parallel\mathbf{a}$, the anomaly at $T_C$ is suppressed at magnetic fields $\mu_0H\geq1$~T. For $\mathbf{H}\parallel\mathbf{b}$ and $\mathbf{H}\parallel\mathbf{c}$, Figures \ref{Fig3}(b-c) show that $T^{max}_{\partial\rho/\partial T}$ is almost unchanged under measured magnetic fields up to 50~T. Figures \ref{Fig3}(b-c) and the focuses shown in Figures \ref{Fig3}(e-f) indicate that, for these two field directions, the anomaly at $T_C$ survives in fields up to 10~T and that it disappears in fields higher than 20~T.

Figures \ref{Fig4}(a-c) show the magnetic-field-temperature phase diagrams extracted from the electrical-resistivity data presented here, for $\mathbf{H}\parallel\mathbf{a}$, $\mathbf{b}$, and $\mathbf{c}$, respectively. They emphasize an anisotropic magnetic-field dependence of the extracted characteristic temperatures. The signature in $\rho_{zz}$ of the transition to the ferromagnetic state at $T_C$ is quickly lost under magnetic fields $\mu_0H\gtrsim1$~T applied along the easy magnetic axis $\mathbf{a}$, at which the magnetic moments are fully polarized \cite{Lhotel2003,Sakon2007} (PPM regime), while it is more robust under magnetic fields applied along the hard magnetic axes $\mathbf{b}$ and $\mathbf{c}$, for which the traces of the transition are lost for $\mu_0H\gtrsim20$~T. Figure \ref{Fig4}(a) shows that, for $\mathbf{H}\parallel\mathbf{a}$, the temperature scales $T_{\Delta\rho_{zz}}^{max}$ and $T_{\partial\rho_{zz}/\partial T}^{max}$, which are characteristic of the broad resistive signal controlled by the electronic correlations, strongly increase while the amplitudes of the associated anomalies in $\Delta\rho_{zz}$ and $\partial\rho_{zz}/\partial T$ decrease. It indicates that the onset of the PPM regime, which is reached when the magnetic moments are polarized, is associated with a progressive field-induced loss of the electronic correlations, whose temperature scale increases with field. Figures \ref{Fig4}(b-c) show that, for $\mathbf{H}\parallel\mathbf{b}$ and $\mathbf{c}$, the temperature scales $T_{\Delta\rho_{zz}}^{max}$ and $T_{\partial\rho_{zz}/\partial T}^{max}$ are nearly field-independent in fields up to 50~T, indicating that the electronic correlations driving the broad contribution to $\rho_{zz}$ are almost unaffected. Interestingly, the temperature scale $T_{\Delta\rho_{zz}}^{max}$ coincides with $T_C$ at zero magnetic field, indicating that both phenomena, i.e., the ferromagnetic ordering and the electronic correlations driving the broad contribution to $\rho_{zz}$, are probably controlled by similar sets of electronic interactions.

\begin{figure}[t]
	\includegraphics[width=\columnwidth]{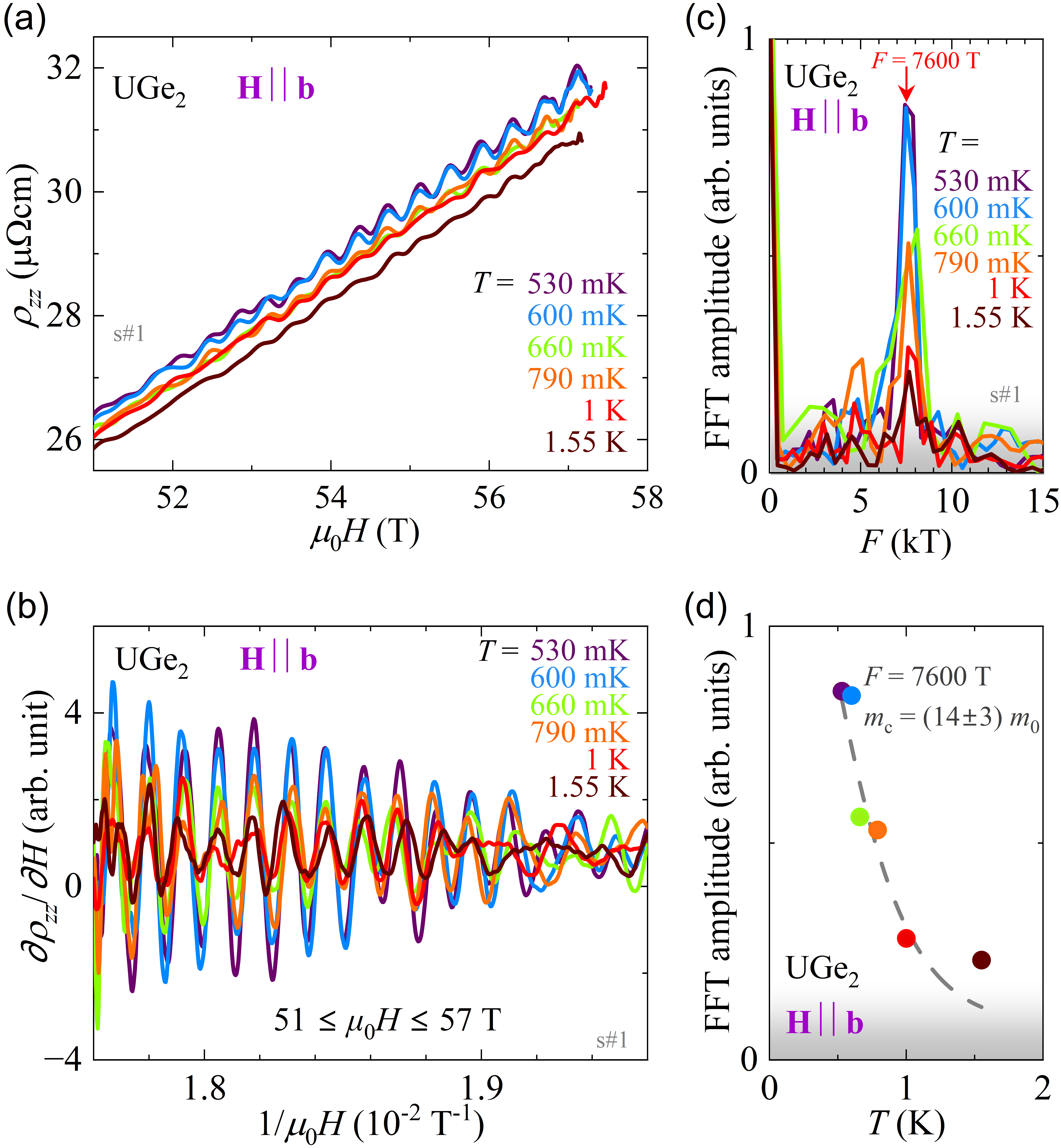}
	\caption{\label{Fig5} Shubnikov-de-Haas quantum oscillations (a) in $\rho_{zz}$ versus $H$ and (b) in $\partial\rho_{zz}/\partial H$ vs $1/H$ of UGe$_2$ at temperatures from 500~mK to 1.55~K under a magnetic field $\mu_0\mathbf{H}\parallel\mathbf{b}$ from 51 to 57.5~T, (c) FFT spectra of $\partial\rho_{zz}/\partial H$ vs $1/H$ at temperatures from 500~mK to 1.55~K and (d) temperature dependence of the amplitude of the peak observed in FFT spectra at the frequency $F=7600$~T and its fit by the Lifshitz-Kosevich formula (dashed grey line).}
\end{figure}

Figure~\ref{Fig5} focuses on the observation of Shubnikov-de Haas quantum oscillations in our electrical resistivity data measured in a magnetic field $\mu_0\mathbf{H}\parallel\mathbf{b}$ larger than 50~T, at temperature ranging from 500~mK to 1.6~K. Figures \ref{Fig5}(a-b) show $\rho_{zz}$ versus $H$ and $\partial\rho_{zz}/\partial H$ vs $1/H$ data emphasizing the presence of the quantum oscillations. The amplitude of the quantum oscillations decreases with increasing temperature and they vanish here for $T>1.55$~K. In Fig.~\ref{Fig5}(c), a fast Fourier transform (FFT) of $\partial\rho_{zz}/\partial H$ vs $1/H$ shows a peak at the frequency $F=7600$~T. According to Onsager, a quantum-oscillation frequency is propositional to the extremal cross section of a Fermi-surface sheet normal to the magnetic field \cite{Onsager1952}. From the temperature dependence of the amplitude of the peak, shown in Figure \ref{Fig5}(d), a cyclotron mass $m_c=(14\pm3)m_0$, where $m_0$ is the free electron mass, is extracted. In the Supplemental Material \cite{SM}, complementary electrical-resistivity measurements show a fast decrease of the intensity of the quantum oscillations when the magnetic field direction is tilted away from $\mathbf{b}$ toward $\mathbf{a}$ or $\mathbf{c}$, which supports that the associated Fermi-surface sheet is cylindrical (see \cite{Onuki1991,Satoh1992_2,Settai2002,Terashima2002,Palacio2016}). These finding are in good agreement with previous reports from de-Haas-van-Alphen experiments, where the main peak from FFT transforms was found with similar frequencies and cyclotron masses \cite{Settai2002,Terashima2002}.

\section{Discussion}
\label{Discussion}

We have studied the electrical resistivity of the prototypical ferromagnet UGe$_2$ under intense magnetic fields applied along its easy and hard magnetic axes. Here, we consider how the electrical resistivity is related with the magnetic anisotropy, but also additional effects visible from our data, as the broad temperature crossover and signatures of the Fermi surface.

Three characteristic temperatures $T_C\simeq52$~K, $T_{\Delta\rho_{zz}}^{max}\simeq50$~K, and  $T_{\partial\rho_{zz}/\partial T}^{max}\simeq30$~K have been extracted. At zero magnetic field, a sharp kink in $\rho_{zz}$, or equivalently a sharp step-like variation in $\partial\rho_{zz}/\partial T$, indicates the transition to ferromagnetic order at $T_C$. A signature of the Ising magnetic anisotropy of UGe$_2$ is that the ferromagnetic transition at $T_C$ disappears under low magnetic fields $>1$~T applied along the easy magnetic axis $\mathbf{a}$, and under much larger fields $\gtrsim20$~T applied along the hard magnetic axes $\mathbf{b}$ and $\mathbf{c}$. A similar anisotropic suppression of ferromagnetic order, due to a faster polarization of the ferromagnetic moments under magnetic field along the easy axis, was also observed in the electrical resistivity of the kagome metal TbV$_6$Sn$_6$ \cite{Rosenberg2022}. A crossover is also in play and the two other temperature scales are related to this crossover, $T_{\Delta\rho_{zz}}^{max}$ at a broad maximum in $\Delta\rho_{zz}$ and $T_{\partial\rho_{zz}/\partial T}^{max}$ at the maximum of $\partial\rho_{zz}/\partial T$. Under magnetic field, this anomaly vanishes faster for $\mathbf{H}$ applied along the easy magnetic axis $\mathbf{a}$ than for $\mathbf{H}$ applied along the hard magnetic axes $\mathbf{b}$ and $\mathbf{c}$, indicating that, as well as ferromagnetism at $T_C$, the correlations driving the crossover also have a magnetic Ising anisotropy.

In heavy-fermion systems, as UGe$_2$ studied here, magnetic fluctuations drive a low-temperature Fermi-liquid contribution $\rho_0+AT^2$ to the electrical resistivity \cite{Knafo2021c}. In UGe$_2$ at temperatures $T\ll T^*$, $A$ can be extracted as function of pressure \cite{Takahashi1993,Tateiwa2001b,Settai2002} or magnetic field (see Supplemental Material \cite{SM}). Under a magnetic field applied along the easy magnetic axis $\mathbf{a}$, the fast decrease of $\rho_{zz}$ is accompanied by a fast collapse of $A$, indicating a fast quenching of the magnetic fluctuations. On the contrary, $\rho_{zz}$ is weakly affected and $A$ is almost unchanged under a magnetic field applied along the hard magnetic axis $\mathbf{c}$. For $\mathbf{H}\parallel\mathbf{b}$, the low-temperature increase of $\rho_{zz}$ induced by the cyclotron motion effect on carriers leads to negative  $\rho_{zz}$ versus $T$ slope a high temperature, ending in non physical negative values of $A$. We speculate that the magnetic fluctuations, which control the enhancement of $A$, are also responsible for the electronic crossover associated with the temperatures $T_{\Delta\rho_{zz}}^{max}$ and $T_{\partial\rho_{zz}/\partial T}^{max}$ in UGe$_2$. The anisotropic response of $\rho_{zz}$ in a magnetic field suggests that these magnetic fluctuations have a Ising uniaxial anisotropy. Longitudinal ferromagnetic fluctuations were indeed identified near to the ferromagnetic transition temperature $T_C$ by inelastic neutron scattering \cite{Huxley2003,Haslbeck2019}. However, no microscopic evidence for low-temperature magnetic fluctuations was reported so far.

In the heavy-fermion paramagnet UTe$_2$, similar broad anomalies were observed in $\rho_{xx}$ and $\rho_{zz}$, measured for $\mathbf{I}\parallel\mathbf{a}$ and $\mathbf{I}\parallel\mathbf{c}$, respectively \cite{Eo2022,Knafo2019,Thebault2022}. The onset of antiferromagnetic fluctuations was suspected to drive these anomalies \cite{Knafo2021b}. Under magnetic field, an anisotropic electrical-resistivity answer, with a faster decrease of $\rho_{xx}$ and $\rho_{zz}$ for $\mathbf{H}$ applied along the magnetic easy axis $\mathbf{a}$ than for $\mathbf{H}$ applied along the hard easy axes $\mathbf{b}$ and $\mathbf{c}$ \cite{Eo2022,Knafo2019,Thebault2022}, suggests that the electronic correlations driving the anomalies in $\rho_{xx}$ and $\rho_{zz}$ have an Ising magnetic anisotropy. An Ising anisotropy of low-temperature magnetic fluctuations was also revealed by NMR relaxation-rate measurements \cite{Tokunaga2019}. For $\mathbf{H}$ applied along the hard magnetic axis $\mathbf{b}$, metamagnetism is induced in UTe$_2$ and leads to non-monotonic variations of $\rho_{xx}$ and $\rho_{zz}$ with $H$. On the contrary, metamagnetism is absent in UGe$_2$ under fields up to 60~T and ambient pressure. Although their electrical resistivity shows similar broad crossovers at zero magnetic field, which are suspected to be induced by magnetic fluctuations, their different high-field responses indicate that different sets of magnetic exchange and anisotropy parameters, and, thus, different magnetic-fluctuations modes, are in play in UTe$_2$ and UGe$_2$.

Finally, carrier cyclotron-motion effects are induced at low temperature and high field for the two transverse configurations with $\mathbf{I}\perp\mathbf{H}$ ($\mathbf{H}\parallel\mathbf{a,b}$), but not for the longitudinal configuration with  $\mathbf{I}\parallel\mathbf{H}\parallel\mathbf{c}$ (see similar effects in URu$_2$Si$_2$ \cite{Scheerer2012}). For $\mathbf{H}\parallel\mathbf{b}$, on top of the cyclotron-motion induced increase of $\rho_{zz}$ we also observed Shubnikov-de-Haas quantum oscillations of $\rho_{zz}$, for which a frequency $F=7600$~T and a cyclotron mass $m_c=(14\pm3)m_0$ have been extracted. Similar frequencies and masses were found from de-Haas-van-Alphen experiments performed at lower magnetic fields $\mu_0H<20$~T and lower temperatures $T\gtrsim30$~mK \cite{Settai2002,Terashima2002}. The high-field increase of $\rho_{zz}$ and its quantum oscillations observed here are probably both controlled by carriers from the same Fermi-surface sheet. The decrease of the absolute value of $\rho_{zz}$ and of the quantum-oscillations amplitude under a magnetic field tilted away from $\mathbf{b}$ (see Supplemental Material \cite{SM}) supports that this Fermi surface is cylindrical. Our measurements also indicate that the Fermi surface of UGe$_2$ is unaffected under a magnetic field of $>50$~T applied along $\mathbf{b}$. This contrasts with the strong pressure-induced modification of the Fermi surface at the pressure $p_x=1.2$~GPa, in the vicinity of which an enhancement of the Fermi liquid coefficients $A$ and $\gamma$ and superconductivity were also reported \cite{Takahashi1993,Tateiwa2001b,Settai2002,Terashima2002}.

\section{Perspectives}

Our motivation was to characterize the electrical resistivity of a typical heavy-fermion U-based ferromagnet under wide ranges of temperature and intense magnetic field applied along different crystallographic directions. Our study of UGe$_2$ showed an anisotropic electrical resistivity in high magnetic field, highlighting the anisotropic and magnetic nature of the electronic correlations. In the future, a magnetization study under high magnetic fields up to 60~T combined with a wide range of temperatures, as done in the resistivity study presented here, could provide complementary information about the anisotropic magnetic properties. In addition, extending resistivity measurements under combined pressure and magnetic field could give further insights about the relation between the magnetic properties and the appearance of superconductivity in UGe$_2$. Given that superconductivity appears near to a pressure-induced magnetic critical point, exploring the signatures of magnetic fluctuations under simultaneous tuning by field and pressure could clarify their role in stabilizing superconductivity.

%\bibliography{UGe2_Kalaiarasan}

\newpage

\onecolumngrid

\renewcommand\thefigure{S\arabic{figure}}
\renewcommand{\theequation}{S\arabic{equation}}
\renewcommand{\thetable}{S\arabic{table}}
\renewcommand{\bibnumfmt}[1]{[S#1]}
\renewcommand{\citenumfont}[1]{S#1}
\setcounter{figure}{0}
\renewcommand{\thesection}{S\arabic{section}}
\renewcommand{\thesubsection}{S\arabic{subsection}}

\vspace{15cm}
\begin{center}
\large {\textbf {Supplemental Material:\\
Anisotropic collapse of electronic correlations in UGe$_2$ under high magnetic field}}
\end{center}
\vspace{1cm}

\setcounter{page}{1}

In this Supplemental Material, we present complementary plots of our electrical-resistivity data. Details about the electrical resistivity $\rho_{zz}$ measured at temperatures near $T_C$ and at the lowest investigated temperature $T=500$~mK, low-temperature Fermi-liquid fits to $\rho_{zz}$ and the extracted magnetic-field variation of the Fermi-liquid coefficient $A$, and the quantum oscillations observed in $\rho_{zz}$ under magnetic fields tilted away from the direction $\mathbf{b}$ are given.

\author{K. Somesh}
\affiliation{Laboratoire National des Champs Magnétiques Intenses- EMFL, CNRS, Univ. Grenoble Alples, INSA-T, Univ. Toulouse 3, 31400 Toulouse, France}
\author{T. Thebault}
\affiliation{Laboratoire National des Champs Magnétiques Intenses- EMFL, CNRS, Univ. Grenoble Alples, INSA-T, Univ. Toulouse 3, 31400 Toulouse, France}
\affiliation{Institute for Quantum Materials and Technologies, Karlsruhe Institute of Technology, D-76021 Karlsruhe, Germany}
\author{V. Taufour}
\affiliation{Univ. Grenoble Alpes, CEA, Grenoble INP, IRIG, PHELIQS, 38000, Grenoble, France}
\affiliation{Department of Physics and Astronomy, University of California, Davis, California 95616, USA}
\author{D. Aoki}
\affiliation{Institute for Materials Research, Tohoku University, Ikaraki 311-1313, Japan}
\author{F. Duc}
\affiliation{Laboratoire National des Champs Magnétiques Intenses- EMFL, CNRS, Univ. Grenoble Alples, INSA-T, Univ. Toulouse 3, 31400 Toulouse, France}
\author{G. Knebel}
\affiliation{Univ. Grenoble Alpes, CEA, Grenoble INP, IRIG, PHELIQS, 38000, Grenoble, France}
\author{D. Braithwaite}
\affiliation{Univ. Grenoble Alpes, CEA, Grenoble INP, IRIG, PHELIQS, 38000, Grenoble, France}
\author{W. Knafo}
\affiliation{Laboratoire National des Champs Magnétiques Intenses- EMFL, CNRS, Univ. Grenoble Alples, INSA-T, Univ. Toulouse 3, 31400 Toulouse, France}

\newpage
\setcounter{page}{1}

\renewcommand\thefigure{S\arabic{figure}}
\renewcommand{\theequation}{S\arabic{equation}}
\renewcommand{\thetable}{S\arabic{table}}
\setcounter{figure}{0}
\onecolumngrid

\begin{figure*}[t]
\includegraphics[width=\textwidth]{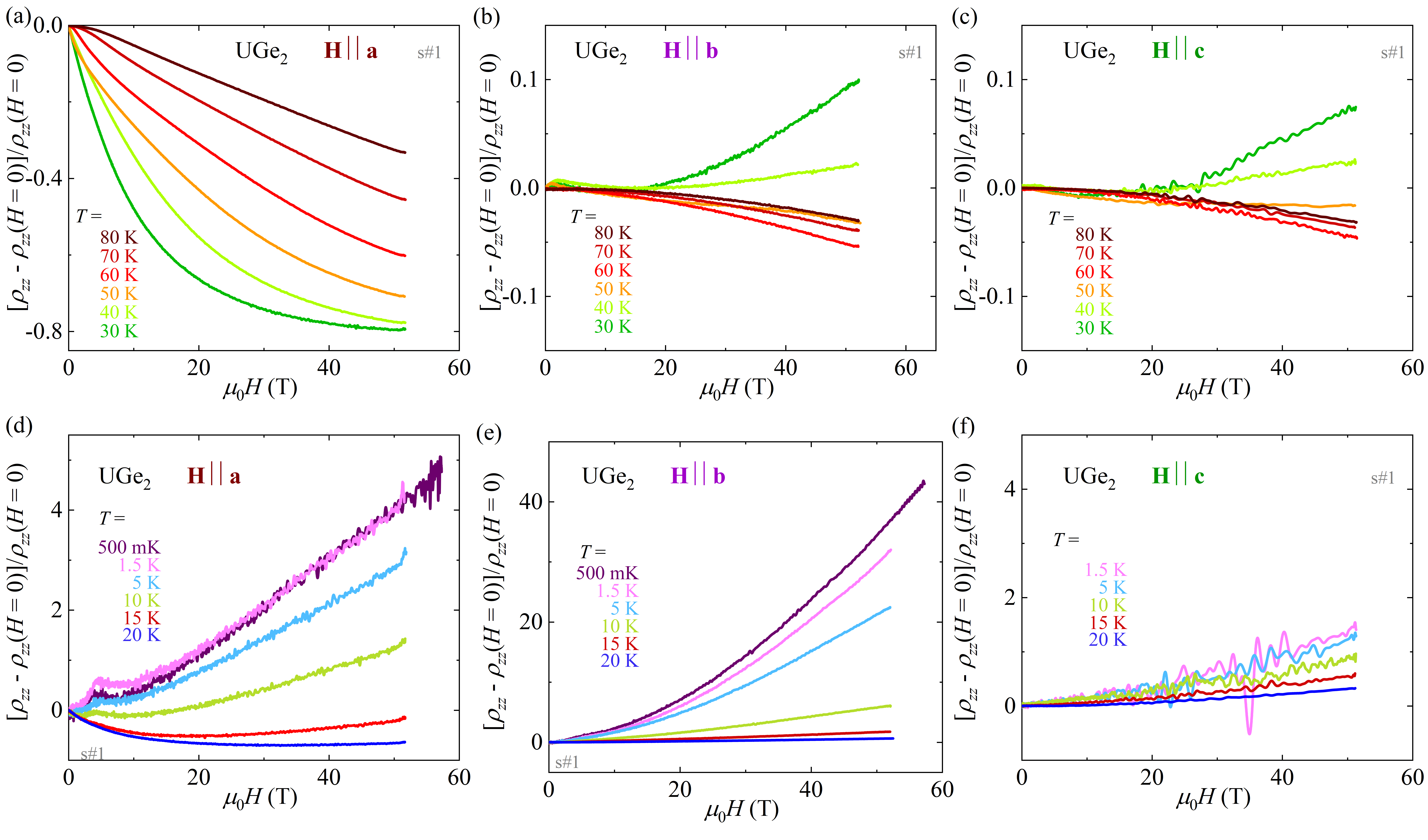}
\caption{\label{FigS1} Magnetic-field dependence of $[\rho_{zz}-\rho_{zz}(H=0)]/\rho_{zz}(H=0)$ of UGe$_2$ at temperatures from 30 to 80~K under magnetic fields (a) $\mu_0\mathbf{H}\parallel\mathbf{a}$, (b) $\mu_0\mathbf{H}\parallel\mathbf{b}$, and (c) $\mu_0\mathbf{H}\parallel\mathbf{c}$ up to 52~T, and at temperatures from 500~mK to 20~K under magnetic fields (d) $\mu_0\mathbf{H}\parallel\mathbf{a}$, (e) $\mu_0\mathbf{H}\parallel\mathbf{b}$, and (f) $\mu_0\mathbf{H}\parallel\mathbf{c}$ up to 52~T}
\end{figure*}

\begin{figure*}[t]
\includegraphics[width=0.34\columnwidth]{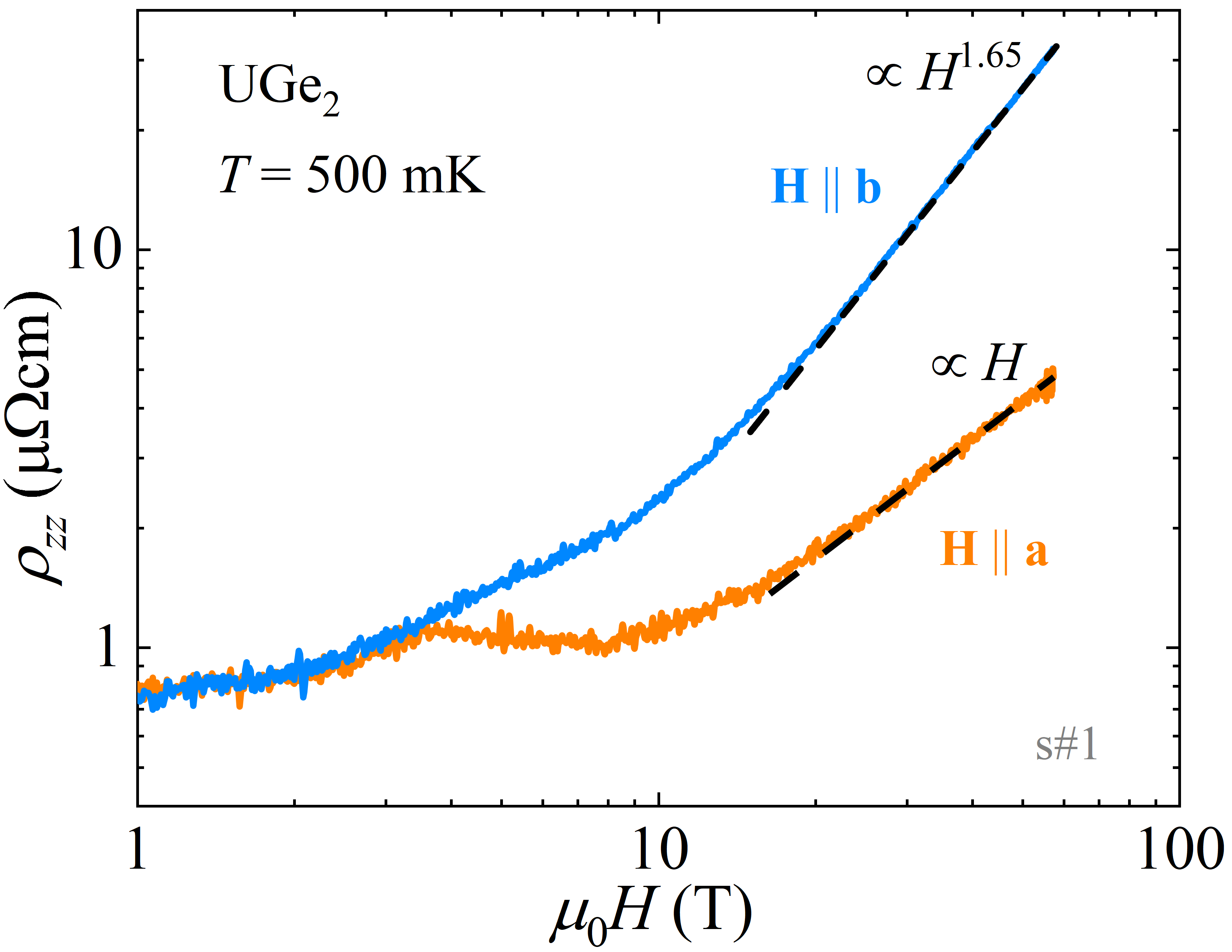}
\caption{\label{FigS2} Magnetic-field dependence of $\rho_{zz}$, in a $log-log$ scale, measured at $T \simeq 500$~mK under magnetic fields $\mu_0\mathbf{H}\parallel\mathbf{a}$ and $\mu_0\mathbf{H}\parallel\mathbf{b}$ up to 58~T. The dashed lines indicate $H$-power-law fits to $\rho_{zz}$ in the limit of high magnetic fields.}
\end{figure*}

Figures \ref{FigS1}(a-c) present plots of the electrical resistivity normalized by the zero-field value $[\rho_{zz}-\rho_{zz}(H=0)]/\rho_{zz}(H=0)$ versus $H$ at temperatures from 30 to 80~K, and Figures \ref{FigS1}(d-f) present plots of $[\rho_{zz}-\rho_{zz}(H=0)]/\rho_{zz}(H=0)$ versus $H$ at temperatures from 500~mK to 20~K, for magnetic fields up to 52~T applied along the three directions $\mathbf{a}$, $\mathbf{b}$, and $\mathbf{c}$, respectively. Alternatively to the plots shown in Figure 1, Figures \ref{FigS1}(a-c) emphasize the larger field variations of $\rho_{zz}$ driven by a faster collapse of electronic correlations under a magnetic applied along the easy magnetic axis $\mathbf{a}$, in temperatures larger than 30~K. They also show that, under a magnetic applied along the hard magnetic axes $\mathbf{b}$ and $\mathbf{c}$, the slope of $\rho_{zz}$ versus $H$ is negative for $T>T_C$ and positive for $T<T_C$. Figures \ref{FigS1}(d-f)  also show that at low temperatures $T\leq20$~K, a field-induced cyclotron-motion-driven enhancement of $\rho_{zz}$ is induced for $\mathbf{H}\parallel\mathbf{a}$ and $\mathbf{H}\parallel\mathbf{b}$, but is almost absent for $\mathbf{H}\parallel\mathbf{c}$. This effect, which is enhanced at high fields and low temperatures, is the strongest for $\mathbf{H}\parallel\mathbf{b}$ due to the presence of cylindrical Fermi-surface sheets along $\mathbf{b}$. Figure \ref{FigS2} shows a plot of the magnetic-field dependence of $\rho_{zz}$, in a $log-log$ scale, measured at $T \simeq 500$~mK under magnetic fields $\mu_0\mathbf{H}\parallel\mathbf{a}$ and $\mu_0\mathbf{H}\parallel\mathbf{b}$ up to 58~T. The increase of $\rho_{zz}$ attributed to a cyclotron-motion effect follows a $H$ law for $\mathbf{H}\parallel\mathbf{a}$ and a $H^{1.65}$ law for $\mathbf{H}\parallel\mathbf{b}$, rather than a $H^2$ law as expected for a two-band compensated-metal model \cite{Onuki2018}.

\begin{figure*}[t]
\includegraphics[width=\textwidth]{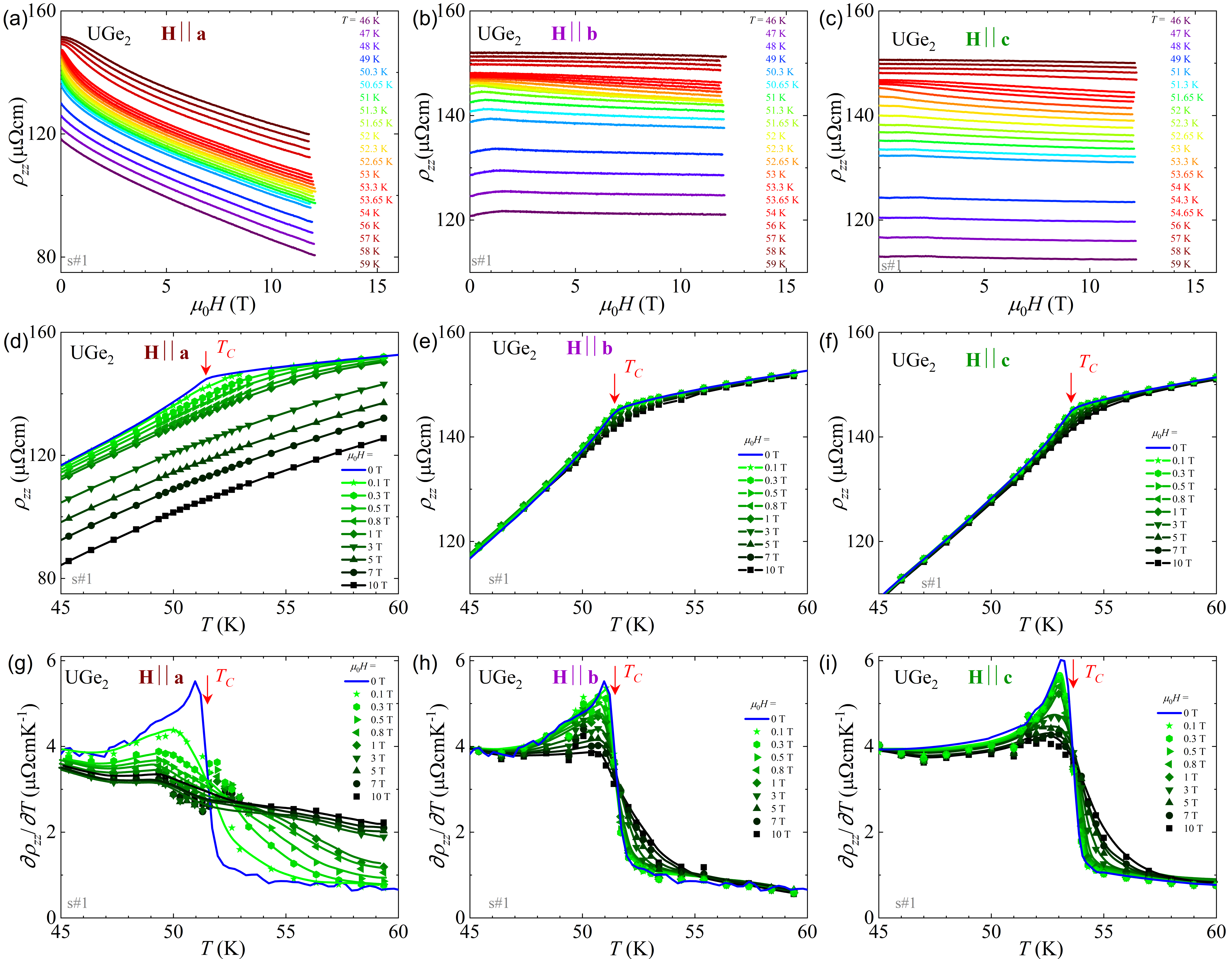}
\caption{\label{FigS3}  Magnetic-field dependence of $\rho_{zz}$ of UGe$_2$ at temperatures from 46 to 59~K under magnetic fields (a) $\mu_0\mathbf{H}\parallel\mathbf{a}$, (b) $\mu_0\mathbf{H}\parallel\mathbf{b}$, and (c) $\mu_0\mathbf{H}\parallel\mathbf{c}$ up to 12~T. Temperature dependence of $\rho_{zz}$ at constant magnetic fields (d) $\mu_0\mathbf{H}\parallel\mathbf{a}$, (e) $\mu_0\mathbf{H}\parallel\mathbf{b}$, and (f) $\mu_0\mathbf{H}\parallel\mathbf{c}$ up to 10~T. Temperature dependence of $\partial\rho_{zz}/\partial T$ extracted from fine $T$-steps data at constant magnetic fields (g) $\mu_0\mathbf{H}\parallel\mathbf{a}$, (h) $\mu_0\mathbf{H}\parallel\mathbf{b}$, and (i) $\mu_0\mathbf{H}\parallel\mathbf{c}$ up to 10~T.}
\end{figure*}

\begin{figure*}[t]
	\includegraphics[width=\textwidth]{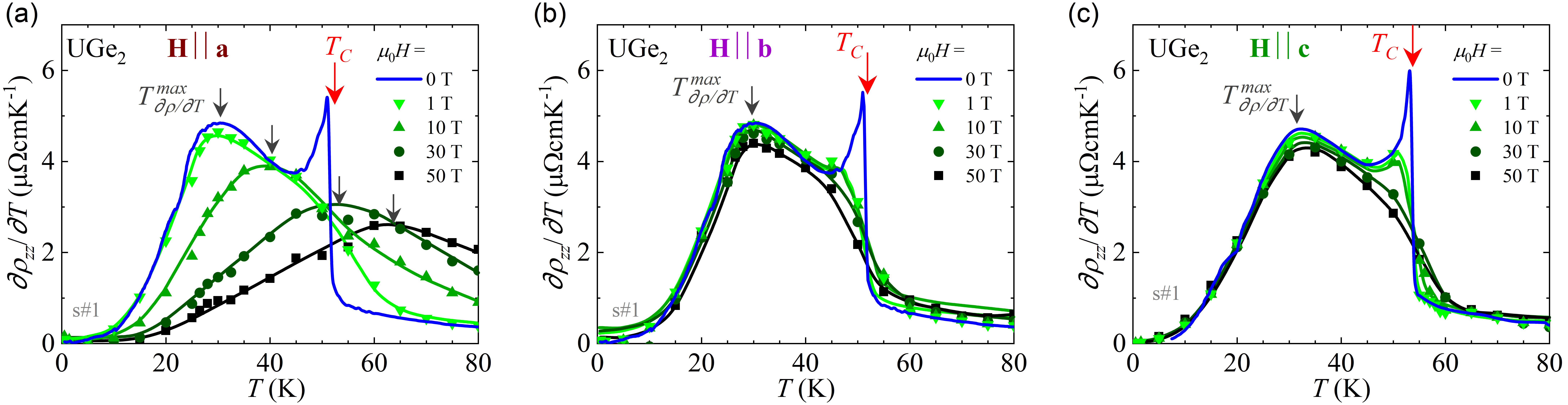}
	\caption{\label{FigS4}  Temperature dependence of $\partial\rho_{zz}/\partial T$ extracted using large $T$-steps data at constant magnetic fields (a) $\mu_0\mathbf{H}\parallel\mathbf{a}$, (b) $\mu_0\mathbf{H}\parallel\mathbf{b}$, and (c) $\mu_0\mathbf{H}\parallel\mathbf{c}$ up to 50~T.}
\end{figure*}

Figures \ref{FigS3}(a-c) present plots of the electrical resistivity $\rho_{zz}$ versus $H$ with fine temperature steps from 46~K to 59~K and magnetic fields up to 12~T, for the three field directions $\mathbf{H}\parallel\mathbf{a}$, $\mathbf{H}\parallel\mathbf{b}$, and $\mathbf{H}\parallel\mathbf{c}$, respectively. A change of curvature of $\rho_{zz}$ versus $H$ is visible in the vicinity of $T_C$. It is more marked for a magnetic field applied along the easy magnetic axis $\mathbf{a}$. Figures \ref{FigS3}(d-f) show the corresponding $\rho_{zz}$ versus $T$ at magnetic fields up to 10~T and the temperature derivatives $\partial\rho_{zz}/\partial T$ extracted from these data are presented in Figures \ref{FigS3}(g-i). Small temperature steps in this series of pulsed-field measurements up to 12~T permitted to carefully study the ferromagnetic transition at the temperature $T_C$ and its disappearance under small magnetic fields. While 12-T pulsed-field shots can be repeated with a fast rate, $>50$-T shots impose to wait one hour between two shots, limiting the number of investigated temperatures. For this reason, a series of small temperature steps could not be done at high magnetic fields, limiting the $T-$resolution of our $\partial\rho_{zz}/\partial T$ curves in magnetic fields $\gtrsim20$~T. However, Figures \ref{FigS4}(a-c) show that $\partial\rho_{zz}/\partial T$ plots determined only from data collected with large $T$ steps are sufficient to extract the anomaly related with the ferromagnetic transition at $T_C$ and to determine at which magnetic field values this anomaly disappears. The comparison between the $\partial\rho_{zz}/\partial T$ plots shown in Figures \ref{FigS3}(a-c) (obtained from fine $T$ steps) and Figures \ref{FigS4}(a-c) (obtained from large $T$ steps) supports that the anomaly at the ferromagnetic transition vanishes in magnetic fields $\mu_0\mathbf{H}\parallel\mathbf{a}$ larger than 1~T and $\mu_0\mathbf{H}\parallel\mathbf{b},\mathbf{c}$ larger than 20~T.

Some of the electrical contacts on the sample were repaired after the first configuration $\mathbf{H}\parallel\mathbf{c}$. For this reason, the $\rho_{zz}$ versus $T$ reference curve measured at zero field for the first configuration $\mathbf{H}\parallel\mathbf{c}$ is not the same that the $\rho_{zz}$ versus $T$ reference curve measured at zero field for the second and third configurations $\mathbf{H}\parallel\mathbf{b}$ and $\mathbf{H}\parallel\mathbf{a}$, respectively. Non-metallic sample holder and probe were used to avoid eddy-currents heating in pulsed magnetic field, leading to thermal gradients between the sample and the thermometers. When the sample was in $^4$He gas at temperatures $T>4.2$K, these gradients led to errors in the estimation of the sample temperature. Small differences between the $\rho_{zz}$ versus $T$ curves measured at zero field for $\mathbf{H}\parallel\mathbf{c}$ and for $\mathbf{H}\parallel\mathbf{b}$ and $\mathbf{H}\parallel\mathbf{a}$ can be seen in Figure \ref{FigS3}. They result from different thermal-gradient conditions.

\begin{figure*}[t]
	\includegraphics[width=\textwidth]{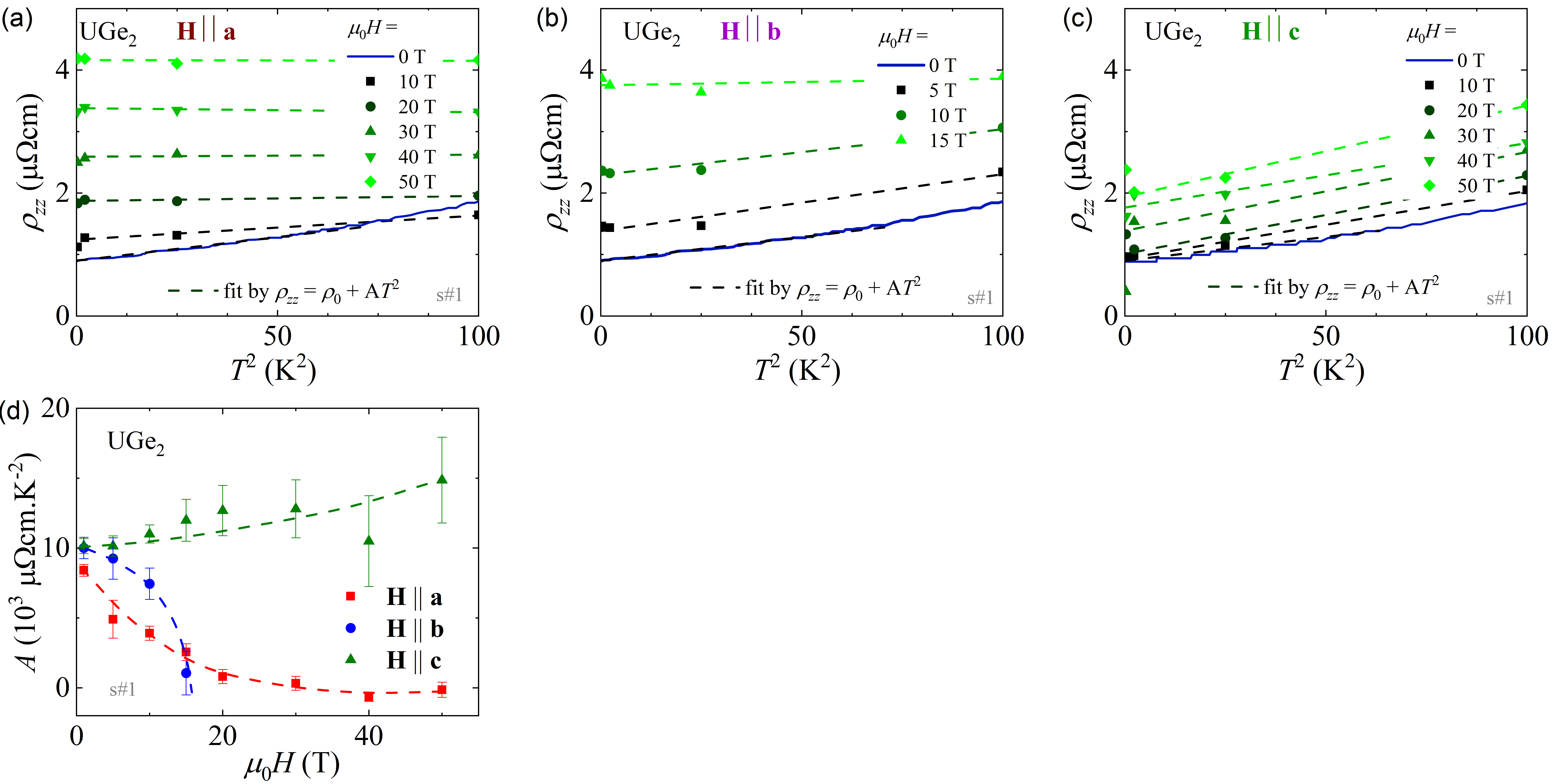}
	\caption{\label{FigS5} $T^2$ variation of $\rho_{zz}$ and Fermi-liquid fits by $\rho_0 + AT^2$, at constant magnetic fields (a) $\mu_0\mathbf{H}\parallel\mathbf{a}$, (b) $\mu_0\mathbf{H}\parallel\mathbf{b}$, and (c) $\mu_0\mathbf{H}\parallel\mathbf{c}$ up to 50~T. (d) Magnetic-field variation of the Fermi-liquid coefficient $A$ for the three magnetic-field directions.}
\end{figure*}

Figures \ref{FigS5}(a-c) show plots of $\rho_{zz}$ versus $T^2$ and Fermi-liquid fits to $\rho_{zz}$ by $\rho_0 + AT^2$, at constant magnetic fields $\mu_0\mathbf{H}\parallel\mathbf{a}$, $\mu_0\mathbf{H}\parallel\mathbf{b}$, and $\mu_0\mathbf{H}\parallel\mathbf{c}$, respectively, up to 50~T. Figure \ref{FigS5}(d) presents the magnetic-field variation of the Fermi-liquid coefficient $A$ extracted from these fits, for the three magnetic-field directions. Under a magnetic field applied along the easy magnetic axis $\mathbf{a}$, a fast collapse of $A$ is found. On the contrary, $A$ is almost unchanged under a magnetic field applied along the hard magnetic axis $\mathbf{c}$. For $\mathbf{H}\parallel\mathbf{b}$, the low-temperature increase of $\rho_{zz}$ controlled by the cyclotron motion effect on carriers leads to negative  $\rho_{zz}$ versus $T$ slope a high temperature, ending in non physical negative values of $A$ at high fields.

\begin{figure*}[t]
	\includegraphics[width=0.5\columnwidth]{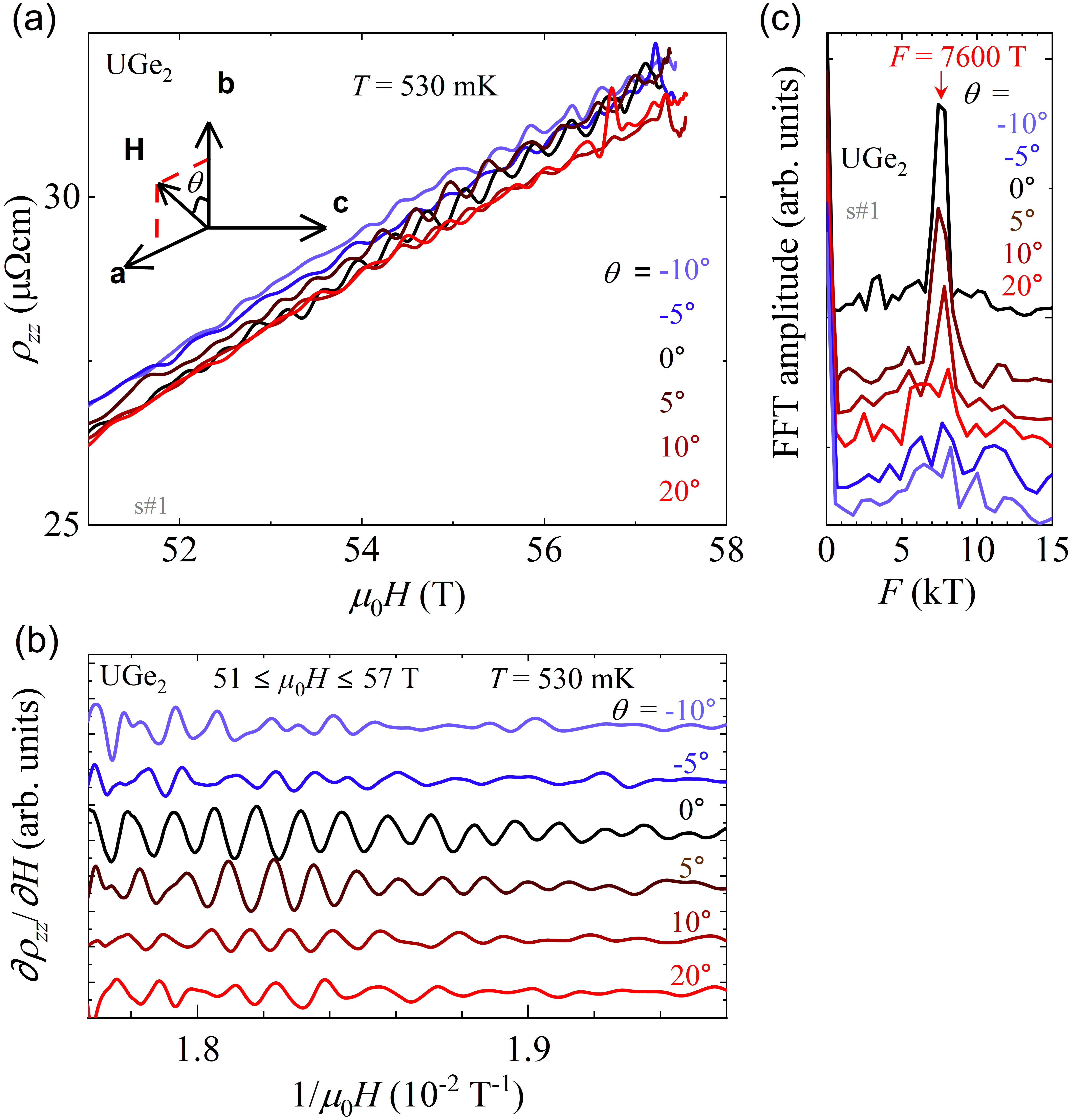}
	\caption{\label{FigS6} Shubnikov-de-Haas quantum oscillations (a) in $\rho_{zz}$ versus $H$ and (b) in $\partial\rho_{zz}/\partial H$ vs $1/H$ of UGe$_2$ at $T=530$~mK under a magnetic field $\mu_0\mathbf{H}$ from 51 to 57.5~T and tilted from $\mathbf{b}$ towards $\mathbf{a}$ by the angle $\theta$ from -10 to 20~$^\circ$. (c) Corresponding FFT spectra of $\partial\rho_{zz}/\partial H$ vs $1/H$.}
\end{figure*}

\begin{figure*}[t]
	\includegraphics[width=0.5\columnwidth]{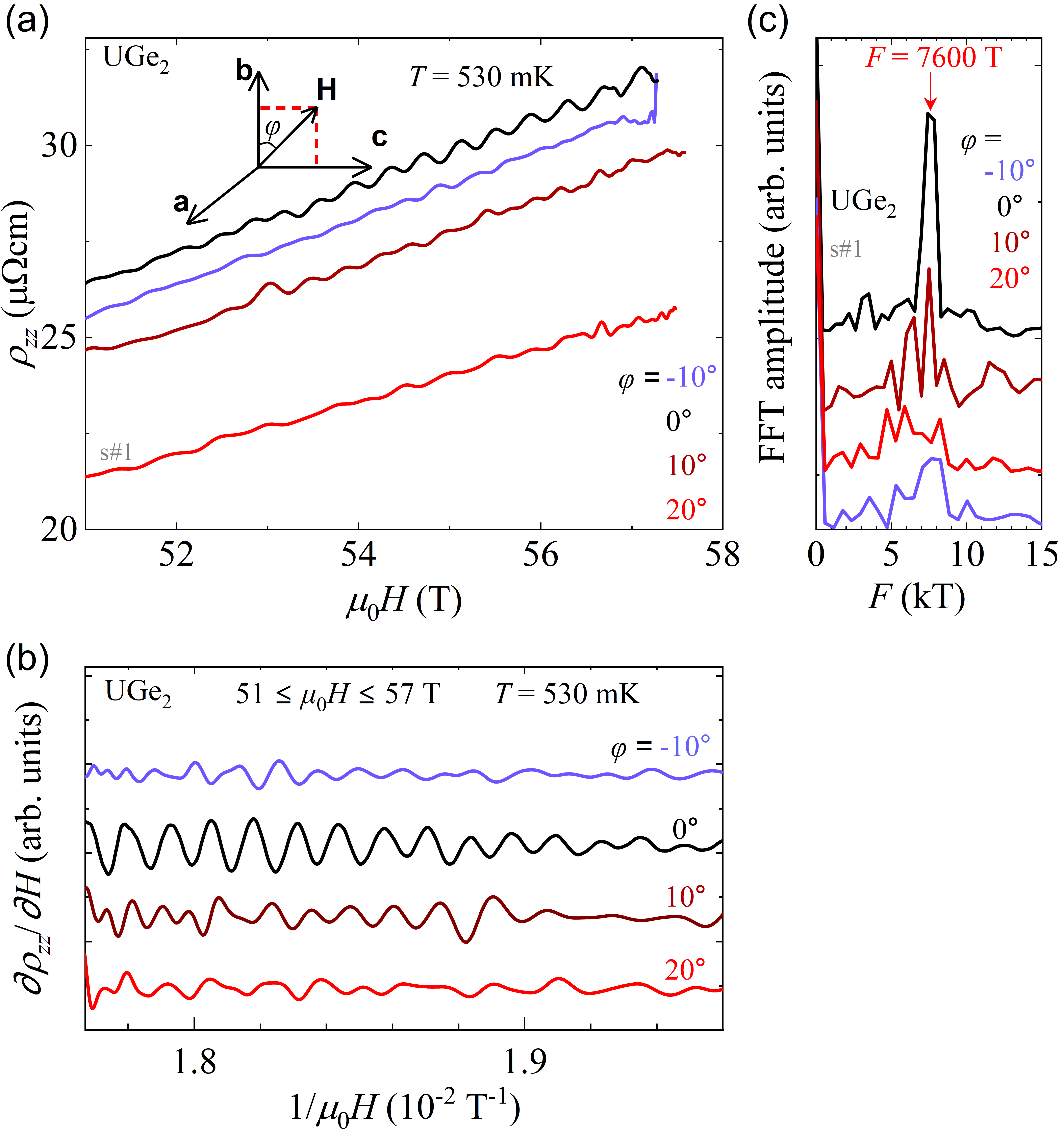}
	\caption{\label{FigS7} Shubnikov-de-Haas quantum oscillations (a) in $\rho_{zz}$ versus $H$ and (b) in $\partial\rho_{zz}/\partial H$ vs $1/H$ of UGe$_2$ at $T=530$~mK under a magnetic field $\mu_0\mathbf{H}$ from 51 to 57.5~T and tilted from $\mathbf{b}$ towards $\mathbf{c}$ by the angle $\varphi$ from -10 to 20~$^\circ$. (c) Corresponding FFT spectra of $\partial\rho_{zz}/\partial H$ vs $1/H$.}
\end{figure*}

Figures \ref{FigS6} and  \ref{FigS7} present details about the Shubnikov-de-Haas quantum oscillations observed in our data, under magnetic fields applied along $\mathbf{b}$ and tilted away from $\mathbf{b}$ towards $\mathbf{a}$ by the angle $\theta$, and towards $\mathbf{c}$ by the angle $\varphi$, respectively. They show that the quantum oscillations at the frequency $F=7600$~T vanish rapidly when the angles $\theta$ and $\varphi$ are increased, supporting that the associated Fermi-surface sheet is a cylinder along $\mathbf{b}$.

%\bibliography{UGe2_Kalaiarasan}

\end{document}